\documentclass[aps,prb,twocolumn]{revtex4}
\usepackage{dcolumn}
\usepackage{bm}
\usepackage{epsfig}
\usepackage{graphicx}
\begin{document}
\title{High energy kink in the dispersion of a hole in an antiferromagnet \\
--- double-occupancy effects on electronic excitations}
\author{Pooja Srivastava, Saptarshi Ghosh, and Avinash Singh}
\email{avinas@iitk.ac.in}
\affiliation{Department of Physics, Indian Institute of Technology Kanpur - 208016}
\begin{abstract}
Evolution of the hole spectral function along the  $\Gamma-(\pi,\pi)$ cut is studied in the antiferromagnetic state of the Hubbard model. The kink in the calculated hole dispersion, the sharp spectral-weight transfer between the branches, 
and the drastically suppressed coherent spectral weight near ${\bf k}=(0,0)$, as observed recently in the high-resolution ARPES studies of cuprate antiferromagnets, are shown to be strongly enhanced by finite-$U$ double-occupancy effects. Together with the anomalous spin-wave dispersion observed earlier in high-resolution neutron-scattering studies, the present study provides further evidence of a unified description of magnetic and electronic excitations in cuprate antiferromagnets in terms of the Hubbard model. 
\end{abstract}
\maketitle
\section{Introduction}
The motion of an added hole/electron in an antiferromagnet (AF) is 
associated with a trail of upturned spins and scrambling of local AF order. 
Theoretical studies of the resulting spin polaron and renormalization of 
quasiparticle properties of doped carriers due to coupling with 
antiferromagnetic spin fluctuations has attracted considerable attention in the context
of spin dynamics, anomalous normal state properties, spin-fluctuation mediated
pairing, optical conductivity etc.\cite{reviews}

Providing important insight into the nature of hole/electron dynamics in an AF, extensive ARPES studies of quasiparticle properties of doped carriers have been carried out recently in cuprates,\cite{arpes} including antiferromagnetic insulators
such as $\rm Sr_2 CuO_2 Cl_2$, $\rm Ca_2 CuO_2 Cl_2$,\cite{wells,kim,ronning_2006} 
and high-temperature superconductors such as $\rm Pb_x Bi_{2-x} Sr_2 CuO_{6+\delta}$, 
$\rm Bi_2 Sr_2 Ca Cu_2 O_{8+\delta}$.\cite{borisenko_2006,graf_2007,xie_2007,valla_2006,pan_2006,chang_2006}
ARPES directly probes the one-particle spectral function $A({\bf k},\omega)$, so it is one of the direct probes of electronic excitations in solids. Over the past decade, a great deal of effort has been invested in further improving this technique. These advances offer unprecedented high momentum and energy resolution, thus making new analysis methods possible and allowing a detailed comprison between theory and experiment.

Most of the theoretical effort to study quasiparticle properties of added hole/electron in the $\rm Cu{O_2}$ plane has concentrated on the $t-J$ model, with double-occupancy excluded completely. Earlier fits of the ARPES data have focussed primarily on the $(\pi,0)-(0,\pi)$ direction along the AF zone boundary, with quasiparticle dispersion calculated in the self-consistent Born approximation (SCBA) for the $t-t'-t''-J$ model, which include second- and third-neighbour hopping terms.\cite{nazarenko,belinicher,kyung,xiang,eder,laughlin,lee,leung} 
The most recent fit\cite{kim} yields parameter values $t=0.35$eV, $t'=-0.12$eV, $t''=0.08$eV, and $J=0.14$eV.

However, with only AF spin interactions included, the Heisenberg model does not provide a complete description of magnetic excitations in cuprate antiferromagnets. High-resolution inelastic neutron scattering studies of the spin-wave spectrum in the cuprate antiferromagnet $\rm La_{2}Cu O_4$ have revealed a noticeable spin-wave dispersion along the AF zone boundary.\cite{neutron} This feature was explained as arising from the higher-order {\em ferromagnetic} spin couplings generated in the strong-coupling expansion for the Hubbard model,\cite{neutron,pallab} including the cyclic ring exchange interaction,
and thus representing finite-$U$, double-occupancy effects on magnetic excitations.

Spectral function and quasiparticle properties have also been obtained recently for the $t-t'-t''$ Hubbard model, with the self energy due to multiple magnon emission/absorption processes evaluated in the noncrossing (rainbow) approximation.\cite{self} Using the same set of Hubbard model parameters as obtained from the spin-wave dispersion fit,\cite{pallab} the quasiparticle dispersion was found to be in good agreement with the ARPES data for $\rm Sr_2 CuO_2 Cl_2$,\cite{wells} thus providing a unified description of both magnetic and electronic excitations in cuprates. Strong finite-$U$ double-occupancy effects due to the large unrenormalized dispersion term $4J\gamma_{\bf k}^2$ were also pointed out on the $\Gamma$ point spectral function, which was shown to have a strongly reduced coherent spectral weight and large spectral-weight transfer to the high-energy incoherent structure. 

Recent high-resolution angle-resolved photoemission spectroscopy (ARPES) experiments on the square-lattice antiferromagnet $\rm Ca_2 CuO_2 Cl_2$,\cite{ronning_2006} with data taken along the $\Gamma-(\pi,\pi)$ cut, have revealed an additional high-energy fast-dispersing feature which merges with the known low-energy feature having strongly renormalized dispersion of $\sim 0.35$. A strong suppression of coherent spectral weight was observed as the $\Gamma$ point is approached, with a rapid spectral-weight transfer from the low- to the high-energy feature. Observed clearly above 0.8 eV binding energy, the fast-dispersing high-energy feature was found to track the unrenormalized band dispersion. 

In this paper we suggest that these novel features observed in the electronic excitations provide further evidence of finite-$U$ double-occupancy effects in cuprate antiferromagnets, already observed in earlier studies of magnetic excitations. 
By examining the evolution of the hole spectral function along the $\Gamma-(\pi,\pi)$ cut, we will show that the large unrenormalized (classical) dispersion term $4J\gamma_{\bf k}^2$ for the hole, arising naturally in the Hubbard model SCBA analysis, has a significant effect on all key features of the hole spectral function near the $\Gamma$ point --- drastically reduced coherent spectral weight, spectral-weight transfer to the high-energy incoherent structure, which tracks the unrenormalized band dispersion. This classical dispersion term is absent in the SCBA studies of the $t-J$ model, as doubly occupied states originally present in the Hubbard model are completely projected out.

The hole self energy for the Hubbard model within the noncrossing (SCBA) approximation is briefly reviewed in section II. Results of our calculation for the hole spectral function are given in section III, showing the evolution along the momentum direction $(\pi/2,\pi/2)-(0,0)$, along with comparisons with the $t-J$ model results and with the
recent ARPES data. Conclusions are presented in section IV.

\section{Hole Self Energy and Spectral function}
The hole self energy in the AF state of the Hubbard model was obtained in the non-crossing (SCBA) approximation as:\cite{self}
\begin{eqnarray}
\Sigma_{\bf k}(\omega) &\equiv &
\langle {\bf k} | [\Sigma({\bf k},\omega)] | {\bf k} \rangle
=(\alpha_{\bf k}^* \;\; \beta_{\bf k}^*)
\left [ \Sigma({\bf k},\omega) \right ]
\left (
\begin{array}{c}
\alpha_{\bf k} \\ \beta_{\bf k} \end{array} \right ) \nonumber \\
&=&
U^2 \sum_{\bf q} \frac{(\alpha_{\bf k} u_{\bf q}\; \beta_{\bf k-q} +
\beta_{\bf k} v_{\bf q}\; \alpha_{\bf k-q} )^2 }
{\omega+ \Omega_{\bf q} -E^\ominus_{\bf k-q} -
\Sigma_{\bf k-q} (\omega+\Omega_{\bf q}) } \; ,
\end{eqnarray}
which represents hole renormalization due to multiple magnon emission and absorption processes. 
Here $\Omega_{\bf q}$ is the RPA-level magnon energy, $u_{\bf q}$, $v_{\bf q}$ are the magnon amplitudes on the two sublattices, and the unrenormalized Hartree-Fock (HF) level hole amplitudes 
$|{\bf k} \rangle \equiv (\alpha_{\bf k} \; \beta_{\bf k})$
and energy $E_{\bf k}^\ominus$ are given by
\begin{eqnarray}
\alpha_{\bf k} ^2 &=& \frac{1}{2}\left ( 1 + \frac{\Delta}
{\sqrt{\Delta^2 + \epsilon_{\bf k} ^2} } \right ) \nonumber \\
\beta_{\bf k} ^2 &=& \frac{1}{2}\left ( 1 - \frac{\Delta}
{\sqrt{\Delta^2 + \epsilon_{\bf k} ^2} } \right ) \nonumber \\
E_{\bf k}^{\ominus} &=& -\sqrt{\Delta^2 + \epsilon_{\bf k}^2 } \nonumber \\
\end{eqnarray}
in terms of free particle energy
$\epsilon_{\bf k} = -2t(\cos k_x + \cos k_y) \equiv -4t\gamma_{\bf k}$,
the antiferromagnetic exchange splitting $2\Delta = mU$, 
and the sublattice magnetization $m$.

In order to make contact with the $t-J$ model result, 
we consider the analytically simple strong-coupling limit $(U/t \gg 1)$, 
where the RPA-level magnon amplitudes and energy are given by\cite{as} 
\begin{eqnarray}
u_{\bf q}^2 &=&
\left (1/\sqrt{1-\gamma_{\bf q}^2} + 1 \right )/2
\nonumber \\
v_{\bf q}^2 &=&
\left (1/\sqrt{1-\gamma_{\bf q}^2} - 1 \right )/2
\nonumber \\
u_{\bf q}v_{\bf q} &=&
\left (-\gamma_{\bf q}/\sqrt{1-\gamma_{\bf q}^2} \right )/2 \nonumber \\
\Omega_{\bf q} &=& 2J \sqrt{1-\gamma_{\bf q}^2} \nonumber \\
\end{eqnarray}
and the hole amplitudes and energy are given by
\begin{eqnarray}
E_{\bf k}^\ominus &\approx& -\Delta - 4J\gamma_{\bf k}^2 \nonumber \\
\alpha_{\bf k} ^2 &\approx& 1 - \epsilon_{\bf k}^2 /U^2 \nonumber \\
\beta_{\bf k} ^2 &\approx& \epsilon_{\bf k}^2 /U^2 \;
\end{eqnarray}
with $J=4t^2/U$. 
Shifting the energy scale $(\omega+\Delta \rightarrow \omega)$ to bring its zero 
at the lower band edge, and with $z$ denoting the lattice coordination number,
the hole self-energy expression (1) reduces to
\begin{equation}
\Sigma_{\bf k}(\omega)
= t^2 z^2
\sum_{\bf q}
\frac{(u_{\bf q} \gamma_{\bf k-q} + v_{\bf q} \gamma_{\bf k} )^2 }
{\omega+\Omega_{\bf q}+4J \gamma_{\bf k-q}^2
- \Sigma_{\bf k-q} (\omega+\Omega_{\bf q}) } \; .
\end{equation}

In comparison with the corresponding $t-J$ model result,\cite{schmitt-rink,gros,kane,liu} 
the above expression for the hole self energy in the Hubbard model involves 
an additional large unrenormalized (classical) dispersion term $4J \gamma_{\bf k}^2$
which has an energy scale twice that of the magnon energy.
The Hubbard and $t-J$ model results therefore differ substantially,
especially near the $\Gamma$ point where the classical band energy 
$4J \gamma_{\bf k}^2 \approx 4J $ has a large contribution in  
the hole spectral function 
\begin{equation}
A_{\bf k}(\omega) = \frac{1}{\pi} {\rm Im} \; 
\frac{1}{\omega + 4 J \gamma_{\bf k}^2 - \Sigma_{\bf k}(\omega)} \; .
\end{equation}

\section{Results}
The self-consistent numerical evaluation of $\Sigma_{\bf k}(\omega)$ was carried out on a 
$52\times 52$ grid in ${\bf k}$ space, and a frequency interval $\Delta 
\omega =0.05$ for $\omega$ in the range $-10 < \omega < 10$, as described earlier.\cite{self}
In the following, we set the energy scale $t=1$. 

\begin{figure}
\vspace*{-02mm}
\hspace*{-05mm}
\psfig{figure=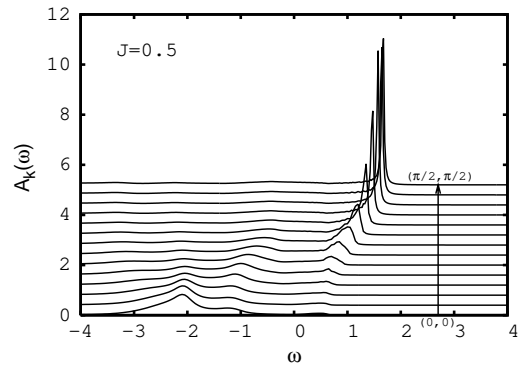,angle=-90,width=70mm}
\vspace{+00mm}
\caption{Evolution of hole spectral function along the $(\pi/2,\pi/2)-(0,0)$
cut for the Hubbard model.}
\end{figure}

Fig. 1 shows the evolution of the hole spectral function for the Hubbard model along the nodal direction ${\bf k}=(\pi/2,\pi/2) - (0,0)$. The low-energy feature forms the well-studied strongly renormalized narrow band, 
with the strongly peaked spectral function near $(\pi/2,\pi/2)$ representing quasiparticles which rapidly lose spectral weight with decreasing $\bf k$. Simultaneously, there is a strong spectral-weight transfer from the quasiparticle to a strongly incoherent high-energy feature which becomes dominant near the $\Gamma$ point. In the following we will highlight the role of the large classical dispersion term in the Hubbard model on this high-energy feature. The energy separation $(\sim 4t)$ between the sharp coherent peak for ${\bf k}=(\pi/2,\pi/2)$ and the broad incoherent peak for ${\bf k}=(0,0)$ is in good agreement with the dynamical cluster quantum Monte Carlo calculations for the Hubbard model.\cite{macridin_2007}

\begin{figure}
\hspace*{-42mm} 
\epsfig{figure=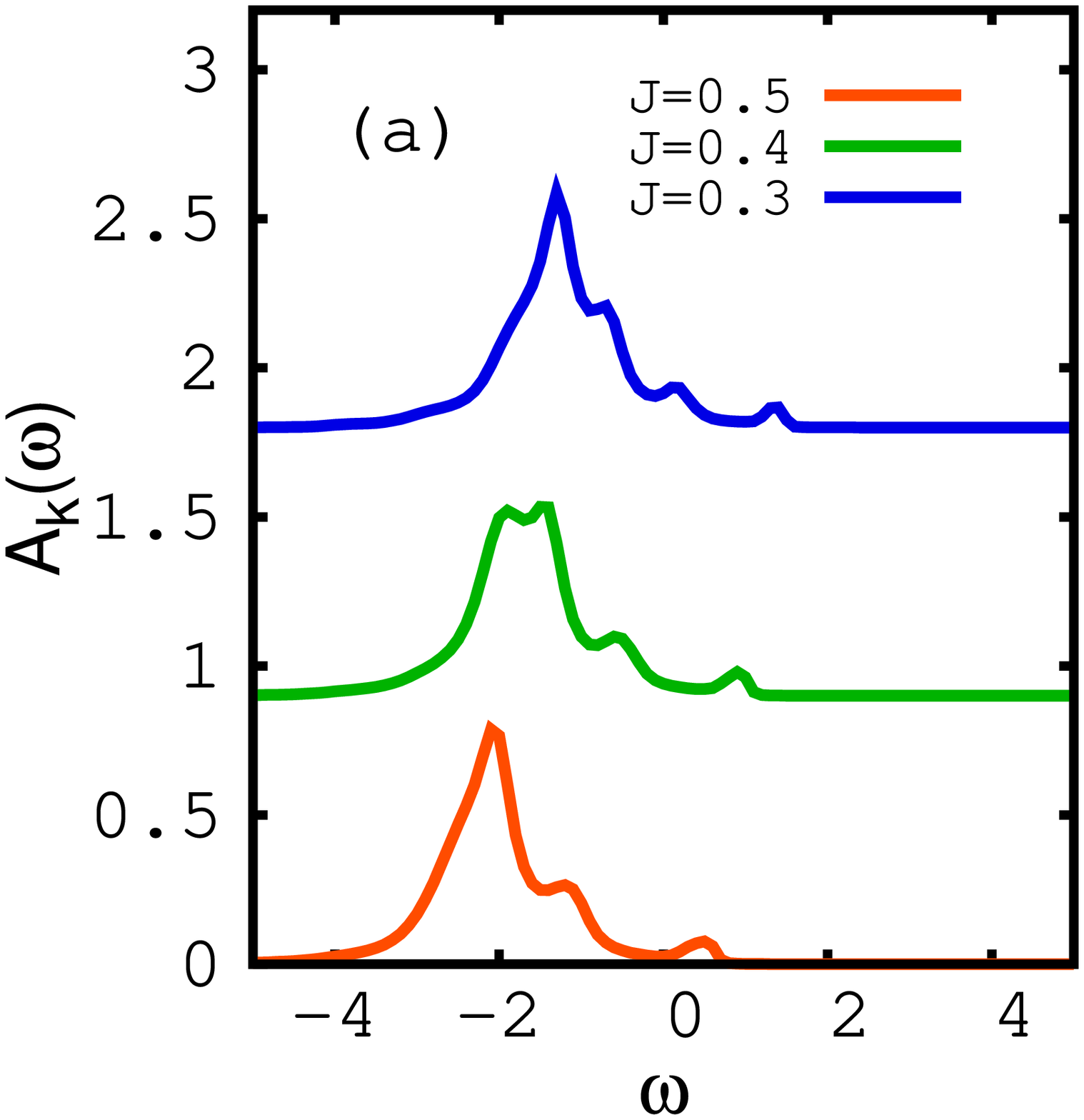,angle=-90,width=45mm}
\end{figure}
\begin{figure}
\vspace*{-50mm} 
\hspace*{43mm}
\epsfig{figure=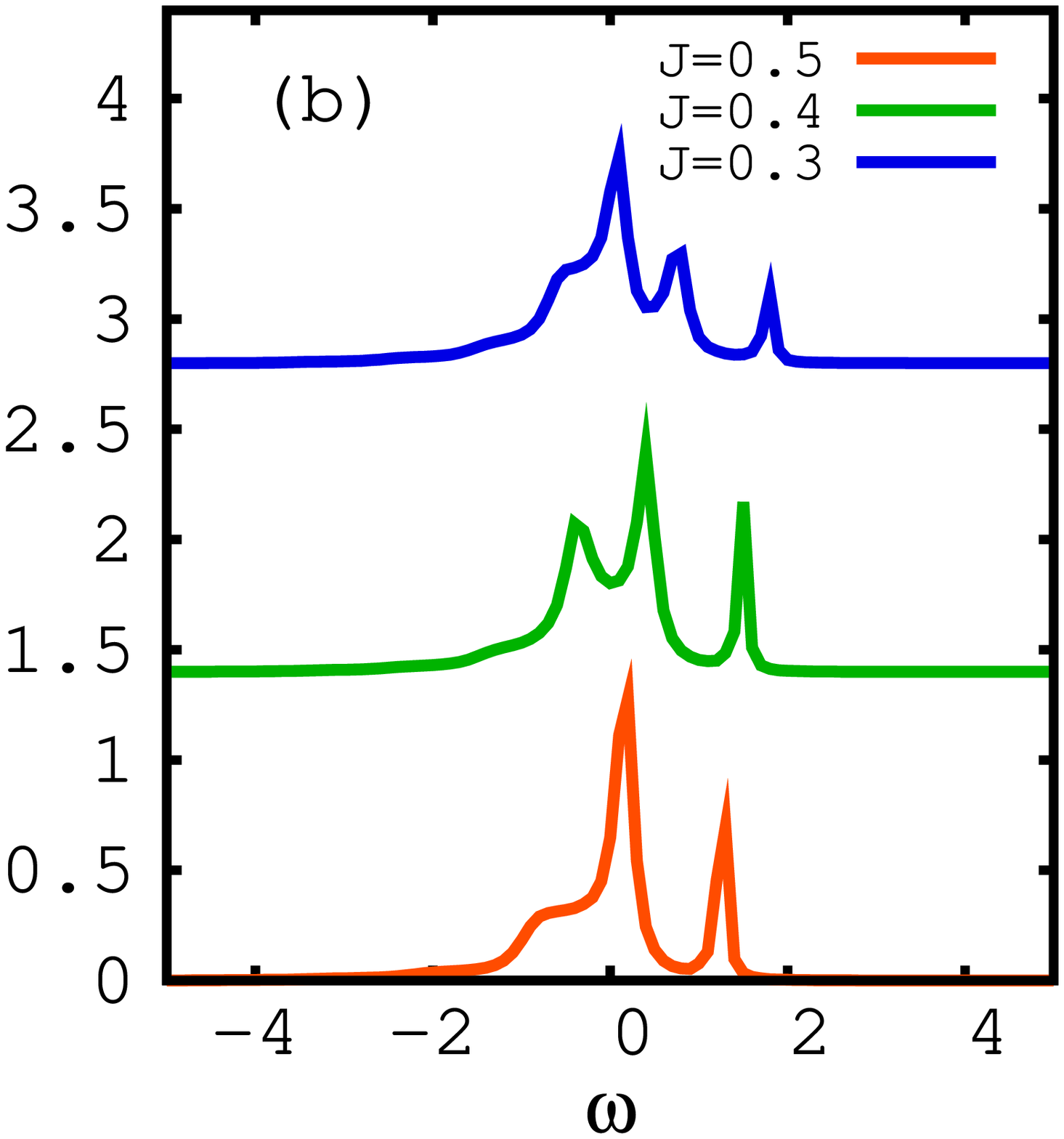,angle=-90,width=45mm} 
\vspace{-00mm}
\caption{Comparison of spectral functions at the $\Gamma$ point 
for the Hubbard model (a) and the $t-J$ model (b).}
\end{figure}

\begin{figure}
\hspace*{-20mm}
\psfig{figure=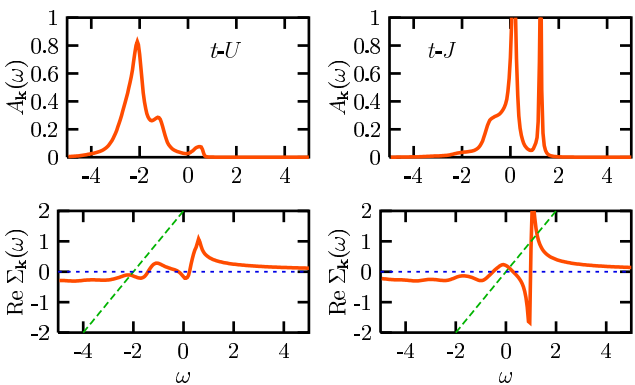,angle=0,width=100mm}
\vspace{-110mm}
\caption{The ${\bf k}=(0,0)$ spectral function is significantly modified by the classical dispersion term, as shown in terms of the intersection of ${\rm Re}\Sigma_{\bf k}(\omega)$ with the lines $\omega+4J\gamma_{\bf k}^2$ and $\omega$ for the Hubbard and $t-J$ models, respectively, with $J=0.5$.}
\end{figure}

Fig. 2 shows a comparison between the Hubbard and $t-J$ model results for the hole spectral function at the $\Gamma$ point for different $J$ values. Presence of the large unrenormalized dispersion term in the spectral function (6) drastically reduces the coherent quasiparticle spectral weight for the Hubbard model, as explained below. Furthermore, this suppression of the coherent spectral weight is a robust feature for the Hubbard model, whereas the overall structure of the spectral function as well as the quasiparticle weight are fairly sensitive to the $J$ value for the $t-J$ model.

Fig. 3, reproduced from Ref. [23], clarifies the role of the large unrenormalized dispersion term $4J \gamma_{\bf k}^2$ on the hole spectral function (6).  For ${\bf k}$ near $(\pi/2,\pi/2)$, the intersection of the line $\omega+4J \gamma_{\bf k}^2$ with ${\rm Re}\Sigma_{\bf k}(\omega)$ occurs near the self energy peak (Fig. 3 in Ref. [23]), resulting in the large hole-energy lowering and the narrow coherent bandwidth. However, for ${\bf k}=(0,0)$, the large classical energy $4J$ shifts the intersection to the high-energy region and away from the self-energy peak, resulting in the strongly reduced coherent quasiparticle weight. Furthermore, the vanishing ${\rm Re}\Sigma_{\bf k}(\omega)$ at the intersection implies that the small-${\bf k}$ states track the unrenormalized band, while the large ${\rm Im}\Sigma$  implies strongly incoherent motion and reflects the strong coupling of the hole with the multiple-magnon string excitations. Furthermore, as the intersection involves the third branch of the self energy (Fig. 3), the high-energy ${\bf k}=(0,0)$ state represents a spin polaron with $\sim $ three magnon excitations.

\begin{figure}
\hspace*{-1mm}
\includegraphics[angle=00,width=0.80\columnwidth]{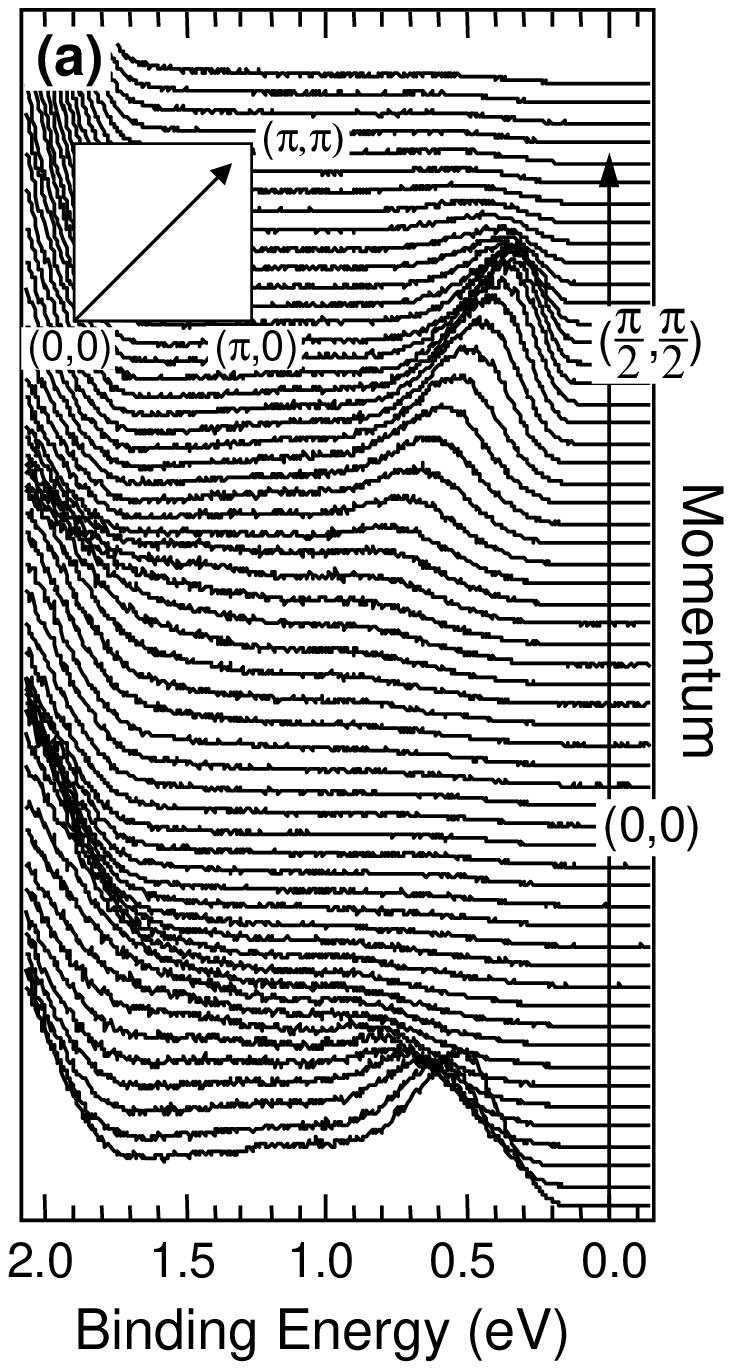}
\hspace*{-35mm}
\includegraphics[angle=90,width=0.50\columnwidth]{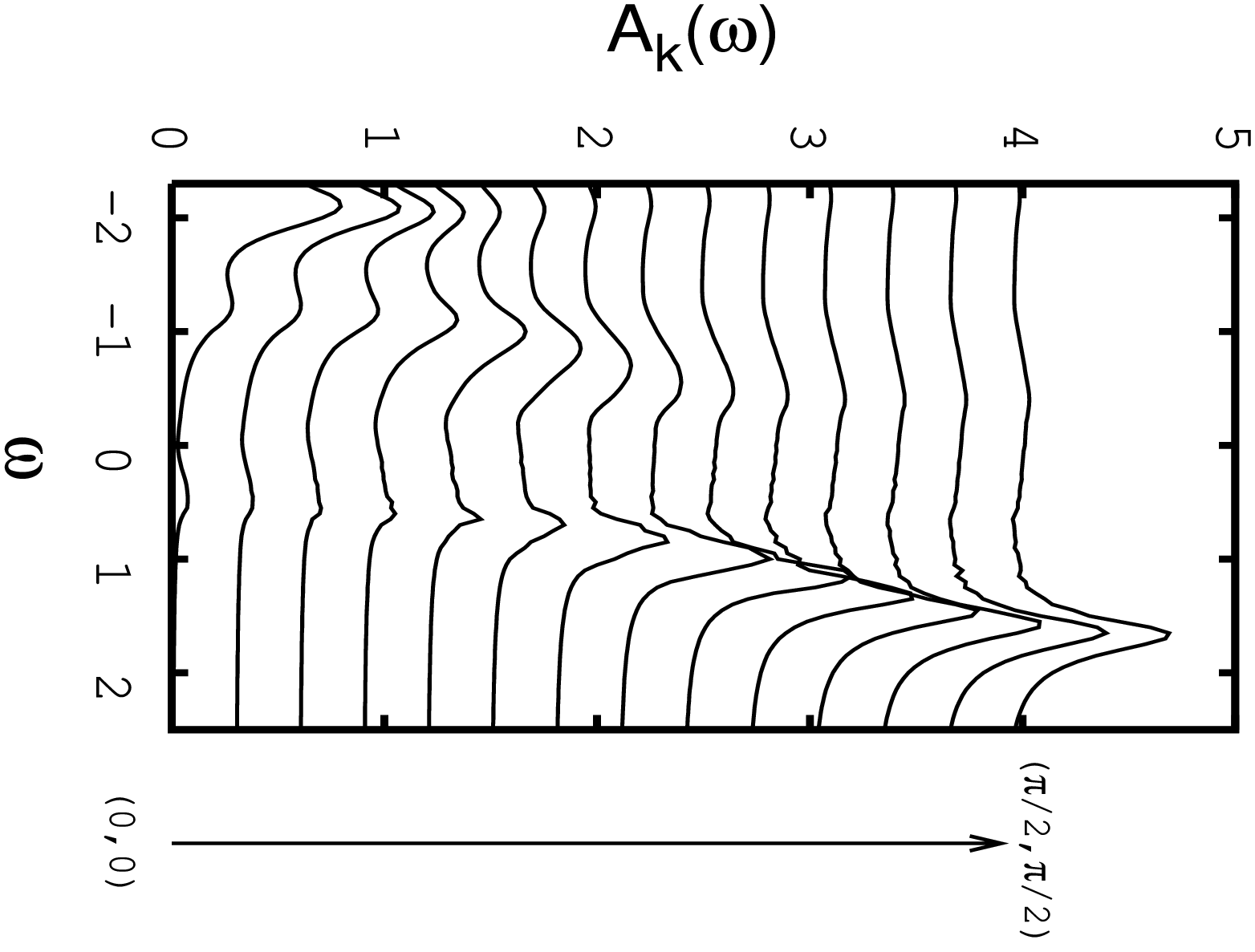}
\\
\caption{Comparison of the hole spectral function along the $(\pi/2,\pi/2)-(0,0)$ direction obtained from the high-resolution ARPES study\cite{ronning_2006} (a) with the Hubbard model result (b).}
\end{figure}

\begin{figure}[ht]
\begin{center}
\includegraphics[angle=00,width=0.75\columnwidth]{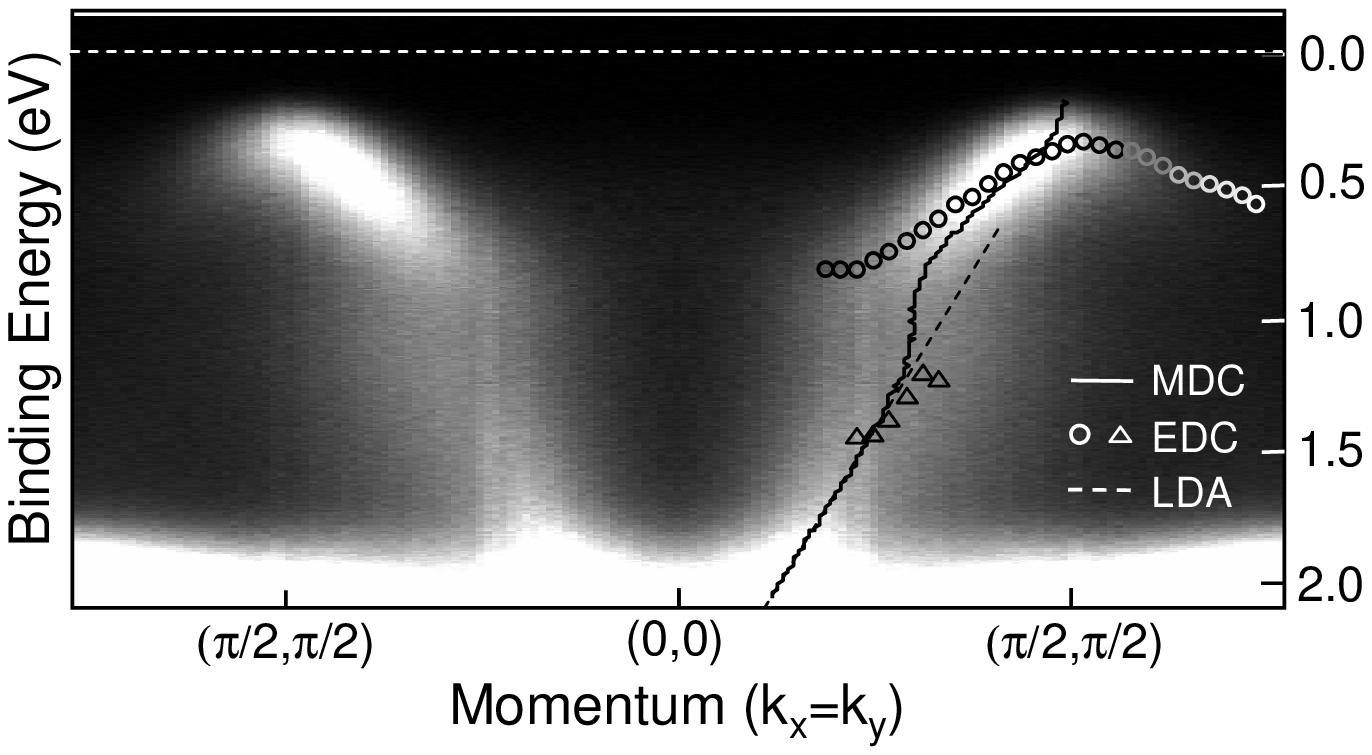} \\
\vspace*{8mm}
\hspace*{-8mm}
\includegraphics[angle=00,width=0.75\columnwidth]{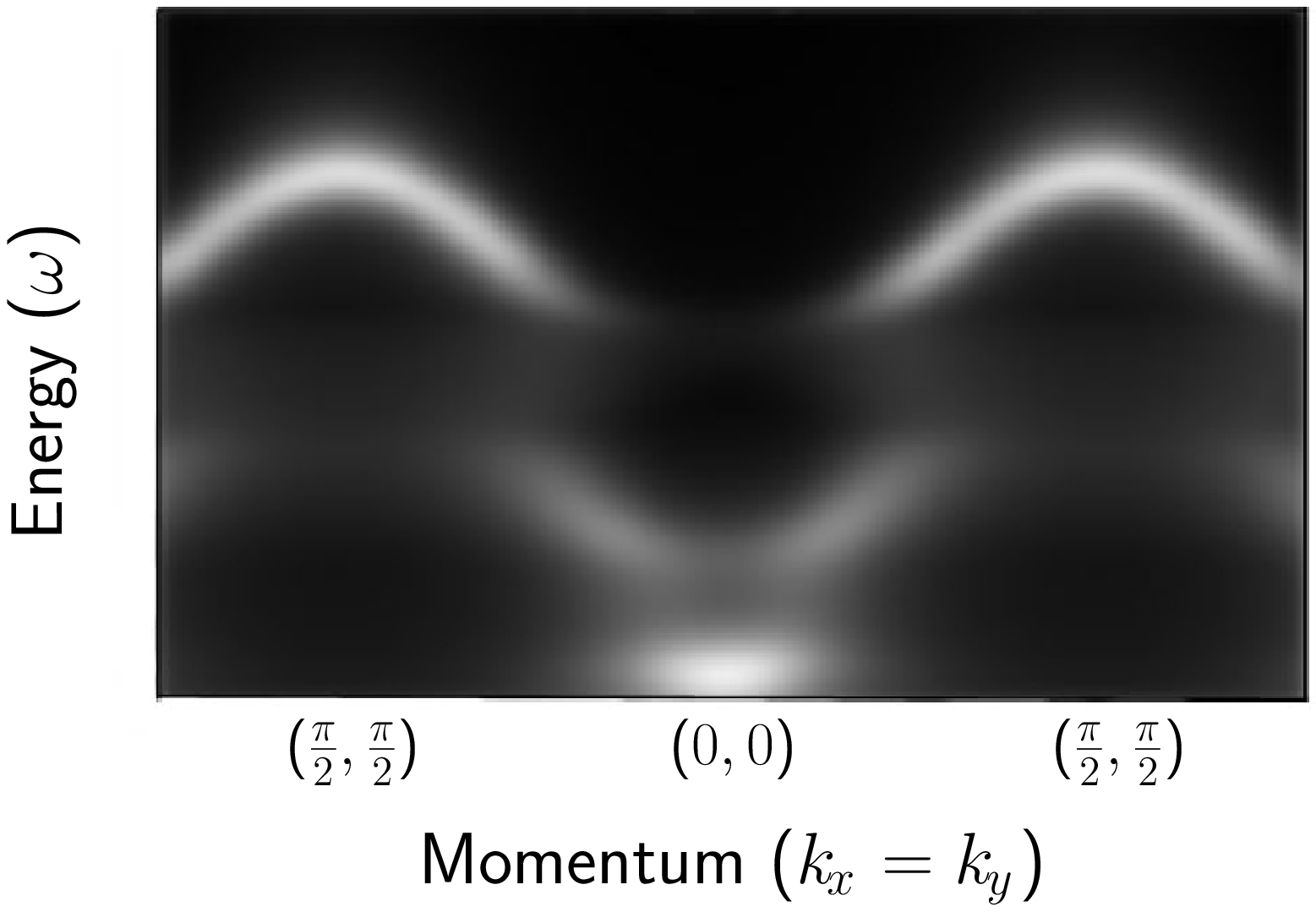}
\vspace*{-0.25cm}
\caption{Comparison of the ARPES intensity plot reported in ref.[5] (upper) and the Hubbard model result (lower), showing the kink in the hole dispersion near $(\pi/4,\pi/4)$. The energy scale is similar for both plots.
The calculated dispersion of both the low- and high-energy features are seen to be in good agreement with the dispersion obtained by following the ARPES peak positions of the MDCs (circles and triangles).}
\end{center}
\label{Fig. 6}
\end{figure}

Fig. 4 shows a comparison of the recent high-resolution ARPES data reported in ref. [5] with the Hubbard model result for $J=0.5$. For ${\bf k}$ near $(\pi/2,\pi/2)$, the spectral function curves in (b) have been 
broadened with a small $\eta$ (Lorentzian broadening) 
in order to bring the spectral function peaks to the same 
scale as in the ARPES data for the purpose of comparison.
No broadening was required for ${\bf k}$ near $(0,0)$ as Im$\Sigma_{\bf k}(\omega)$ is already substantial.
The vanishing spectral weight for ${\bf k}=(0,0)$ observed in photoemission 
experiments on cuprates finds a natural explanation 
in terms of the classical dispersion due to double occupancy within the Hubbard model.
Highlighting the presence of string-like excitations in the AF,
detailed comparison of the ARPES data with the $t-J$ model results has also been recently reported,\cite{manousakis}
where a Gaussian broadening was introduced over the whole $\bf k$ range. 

The inability of the SCBA calculations to reproduce properly the finite linewidth seen in the ARPES plots for ${\bf k}$ near $(\pi/2,\pi/2)$ has been pointed out.\cite{ronning_2006} Since magnon states are not exact eigenstates of the AF, a finite magnon damping must clearly be included, especially for the short wavelength, high-energy modes. 
                                                                           
Fig. 5 shows a comparison of the ARPES intensity plot (upper) with the Hubbard model result (lower) of Fig. 4.
With $t\sim 0.35$eV, the energy scale in the calculated plot ($5t$) is similar as in the ARPES plot.
Key features of the ARPES intensity plot ---
sharp spectral weight transfer and the kinklike feature (``waterfall'') between the 
weakly dispersing low-energy feature near $(\pi/2,\pi/2)$
and the rapidly dispersing high-energy feature near $(0,0)$ ---
are in strong resemblance with the calculated result.
Also, the energy separation $(\sim 4t = 1.4{\rm eV})$ between the sharp coherent peak for ${\bf k}=(\pi/2,\pi/2)$ and the broad incoherent peak for ${\bf k}=(0,0)$ is in good agreement with the ARPES binding energy scale.
Furthermore, the calculated dispersion of both low- and high-energy features are in good agreement with the dispersion obtained by following the ARPES peak positions of the MDCs (circles and triangles).
The energy separation of about $\sim 1.3t \approx 0.4{\rm eV}$ associated with the kink in the dispersion near $(\pi/4,\pi/4)$ is also in good agreement with the separation between the two distinctive high energy scales ($E_1=0.38$ eV and $E_2=0.8$ eV) reported for the high-temperature superconductors.\cite{graf_2007}

\begin{figure}
\vspace*{5mm}
\hspace*{-20mm}
\psfig{figure=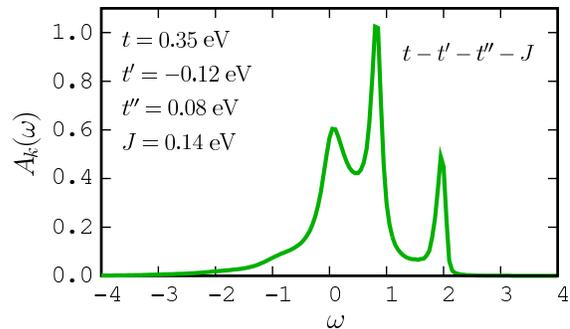,width=90mm}
\vspace{-100mm}
\caption{The calculated $\Gamma$ point spectral function for the $t-t'-t''-J$ model
with parameters obtained in Ref. 4 by fitting the quasiparticle dispersion along the MBZ boundary $(\pi,0)-(0,\pi)$.}
\end{figure}

\begin{figure}
\vspace*{5mm}
\hspace*{-33mm}
\psfig{figure=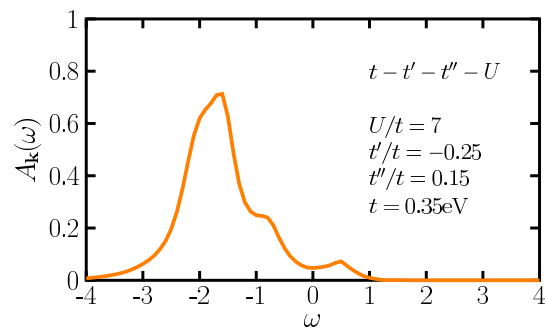,width=100mm}
\vspace{-60mm}
\caption{The calculated $\Gamma$ point spectral function for the Hubbard model
with parameters obtained in Refs. [21,22] by fitting both the 
spin-wave and quasiparticle dispersions along the MBZ boundary.}
\end{figure}

Turning to realistic cuprate parameters,
Fig. 6 shows the calculated $\Gamma$ point spectral function for the $t-t'-t''-J$ model
with best-fit parameters ($J/t=0.4,\; t'/t=-1/3,\; t''/t=1/4.4$) obtained in Ref. 4  
by fitting the quasiparticle dispersion along the MBZ boundary $(\pi,0)-(0,\pi)$. 
As seen, the $t-t'-t''-J$ model result also yields fairly sharp coherent spectral function peak.
For comparison, Fig. 7 shows the calculated $\Gamma$ point spectral function for the Hubbard model
with best-fit parameters ($U/t=7$, $t'/t=-0.25$, $t''/t=0.15$, $t=0.35$eV) 
obtained by fitting both the spin-wave as well as quasiparticle dispersions along the MBZ boundary.\cite{pallab,self}
It is clear that the Hubbard model result yields a much better agreement with respect to the strongly reduced quasiparticle spectral weight observed in ARPES studies.

\section{Conclusions}
In the context of recent high-resolution ARPES experiments,
we studied the evolution of the hole spectral function in the $(0,0)-(\pi,\pi)$ direction
and examined the role of the large unrenormalized (classical) band dispersion term $4J\gamma_{\bf k}^2$ in the Hubbard model

For ${\bf k}$ approaching $(0,0)$,
the large classical band energy ($\sim 4J$) was shown to result in a strong suppression of the coherent quasiparticle spectral weight in the low-energy, strongly renormalized band,
and a rapid transfer of spectral weight to a high-energy incoherent structure which tracks the unrenormalized band dispersion. While for ${\bf k} \sim (\pi/2,\pi/2)$, the hole propagates coherently in the AF background, with a large quantum energy lowering associated mainly with single magnon emission-absorption process, the high-energy incoherent ${\bf k} \sim (0,0)$ hole state was found to correspond to a spin polaron with multiple ($\sim$ three) magnon (string) excitations. 

For ${\bf k}=(0,0)$, the overall structure of the hole spectral function as well as the insignificant coherent spectral weight were found to be quite robust with respect to the magnitude of $J$ for the Hubbard model, whereas the $t-J$ model results exhibited pronounced sensitivity. The calculated hole spectral function for the $t-t'-t''-J$ model,
with same parameters as obtained from the quasiparticle dispersion fit along the MBZ zone boundary in the $(\pi,0)-(0,\pi)$ direction, yielded fairly sharp coherent spectral function peak and the overall shape was also not in agreement with ARPES. 

The large classical dispersion term in the Hubbard model clearly provided a better description of both the strongly suppressed coherent spectral weight observed in ARPES studies and of the high-energy incoherent peak tracking the unrenormalized band dispersion, thus highlighting the importance of finite-$U$, double-occupancy effects on electronic excitations in cuprate antiferromagnets. In view of the observed spin-wave dispersion along the MBZ zone boundary in high-resolution neutron-scattering studies, and their understanding in terms of finite-$U$, double-occupancy effects, 
the present study provides further evidence of a unified description of both magnetic and electronic excitations in cuprate antiferromagnets within an effective Hubbard model description. 

As the dominant contribution to hole/electron self energy due to multiple magnon emission/absorption processes involves the short-wavelength modes which only require short-range AF order, extension of this study to finite doping is of considerable interest. The high-energy feature characterized by sharp spectral-weight transfer and kink in the dispersion has been observed to be a universal anomaly in recent ARPES studies on doped cuprates, independent of doping as well as chemical composition.\cite{graf_2007,xie_2007,valla_2006,pan_2006,chang_2006}

\end{document}